\documentclass[journal]{IEEEtran}

\IEEEoverridecommandlockouts

\usepackage{amsmath,amssymb,amsfonts,empheq,bm}
\usepackage{algorithmic}
\usepackage{graphicx}
\usepackage{textcomp}
\usepackage{xcolor}
\usepackage{float}
\usepackage{setspace}
\usepackage{comment}
\usepackage{cleveref}
\usepackage{dotlessi}
\usepackage{mathtools}
\usepackage{subcaption}
\usepackage[font=footnotesize]{caption}
\usepackage{multirow}
\usepackage{amsmath}
\usepackage{amsfonts}
\usepackage{bm}
\usepackage{etoolbox}
\usepackage{cite}
\usepackage{csquotes}

\makeatletter
\patchcmd{\@sect}{\MakeLowercase}{}{}{}
\makeatother

\newcommand{\jj}{\mathrm{j}}

\newcommand{\teoR}{\Psi_{\mathbb{R}}}
\newcommand{\teoB}{\Psi_{\mathbb{B}}}
\newcommand{\teoC}{\Psi_{\mathbb{C}}}

\begin{document}

\title{Instantaneous Frequency in Power Systems using the Teager-Kaiser Energy Operator} 

\author{%
  {\'A}ngel Vaca, \IEEEmembership{IEEE Student Member}, %
  Joan Guti{\'e}rrez-Florensa, \IEEEmembership{IEEE Student Member}, %
  \\and Federico Milano, \IEEEmembership{IEEE Fellow}%
  \thanks{The authors are with the School of Electrical and Electronic Engineering, University College Dublin, Belfield Campus, D04V1W8, Ireland.  e-mails: angel.vaca1@ucdconnect.ie, joan.gutierrezflorensa1@ucdconnect.ie, federico.milano@ucd.ie}%
  \thanks{This work is supported by the Science Foundation Ireland (SFI) by funding the authors under NexSys project, Grant No.~21/SPP/3756.}
\vspace{-0.7cm}
}%

\maketitle

\begin{abstract}
  This letter develops an instantaneous-frequency (IF) local estimator calculated with the complex Teager--Kaiser energy operator (CTKEO) and the dynamic-signal identity.  The contribution is a novel CTKEO-based IF expression that makes the envelope-curvature terms explicit, thus correcting the bias that affects conventional estimators used in power systems.  The estimator aligns with complex-frequency (CF) kinematics and admits a geometric interpretation (curvature) without phase unwrapping.  This yields an accurate local frequency estimate in operating conditions where magnitude variations contribute non-negligibly to the signal dynamics.  Tests on field measurements illustrate the practical behavior of the proposed approach and its consistency with a geometric-frequency benchmark.
\end{abstract}

\begin{IEEEkeywords}
  Teager-Kaiser Energy Operator (TKEO), instantaneous frequency, complex frequency, differential geometry.
   \vspace{-0.25cm}
\end{IEEEkeywords}

\vspace{-0.35cm}

\section{Introduction}

The estimation of instantaneous frequency (IF) remains a central problem in signal analysis, fundamental to characterizing nonstationary and time-varying phenomena \cite{mandel_interpretation_1974, cohen1995time, 11177640}. In recent years, power-system frequency has also been interpreted through complex-frequency kinematics and the geometry of voltage trajectories \cite{9917164, MilanoFrenet}.

In this work, we consider a different approach to estimate IF based on the Teager-Kaiser Energy Operator (TKEO). This operator provides a local relation between signal derivatives and instantaneous energy \cite{115702, BOUDRAA2018338, Vakman1996OnTA}.  Conventional TKEO-based estimators, however, rely on narrow-band and slow-envelope assumptions, which may lead to biased estimates during transients or when voltage magnitudes vary significantly.  In this context, ``slow'' means that the envelope and frequency-modulation rates are small with respect to the carrier angular frequency, e.g., $|a'|/|a| \ll |\omega|$ and the characteristic bandwidth of $\omega(t)$ is much smaller than the carrier frequency. 

This letter extends the TKEO to complex signals and derives a formulation that unifies nonlinear energy and geometric interpretations.
In particular, the proposed formulation makes explicit the correction terms associated with voltage-magnitude dynamics, which are neglected in simplified estimators and become relevant under nonstationary operating conditions. The resulting estimator is consistent with complex-frequency kinematics and admits a geometric interpretation without requiring explicit phase unwrapping.  A simplified approximation is also discussed for conditions in which magnitude variations are slow.

The proposed method provides a CTKEO-based local realization of the angular component of the complex frequency, consistent with existing complex-frequency and geometric-frequency interpretations.  The novelty of the proposed formulation is that it expresses this angular frequency through the CTKEO while retaining the magnitude-dynamics correction term. This explicitly shows the approximation introduced when such a term is neglected.

\section{Complex Teager--Kaiser Energy Operator}
\label{sec:ESA}

The conventional continuous-time Teager--Kaiser energy operator for real signals is defined as:
\begin{equation}
\teoR(x) = (x')^2 - x\,x''\,.
\label{eq:TKEO-real}
\end{equation}
This quantity measures the instantaneous ``frequency-weighted energy'' of $x$.  
In discrete time, for a time series $x_1, x_2, \dots, x_N$, it takes the form:
\begin{equation}
\teoR(x_n)=(x_n)^2 - x_{n-1}x_{n+1}\, , \quad n \in \{2, \dots, N-1\}.
\label{eq:TKEO-discrete}
\end{equation}

An established application of the Teager--Kaiser operator is the Energy Separation Algorithm (ESA), originally developed for the demodulation of amplitude--frequency modulated (AM--FM) signals \cite{Maragos19933024}.  Consider a real AM--FM component:
\begin{equation}
x = a\cos(\phi)
       = a\cos\!\left(\int_{0}^{t}\omega\,d\tau + \phi(0)\right)\,,
\label{eq:x_t}
\end{equation}
where both $a$ and $\omega = \phi'$ vary slowly relative to the carrier.  When the real TKEO defined in~\eqref{eq:TKEO-real} is applied to such a signal, it yields:
\begin{equation}
\teoR(x) \approx a^2\omega^2\,,
\label{eq:psi_x}
\end{equation}
which represents an instantaneous, frequency-weighted energy proportional to the square of both amplitude and angular frequency.  This approximation holds under the so-called ''narrow-band'' or ``slow-envelope'' condition, and is verified in \cite{115702} and others.

Applying the operator to the time derivative of the signal produces:
\begin{equation}
\teoR(x') \approx a^2\omega^4\,,
\label{eq:psi_xdot}
\end{equation}
and combining~\eqref{eq:psi_x} and~\eqref{eq:psi_xdot} eliminates the amplitude term, giving the classical ESA relations:
\begin{equation}
a\approx \frac{\teoR(x)}{\sqrt{\teoR(x')}}\,, \qquad
\omega \approx \sqrt{\frac{\teoR(x')}{\teoR(x)}}
\label{eq:f_t}
\end{equation}

Equations~\eqref{eq:f_t} provide instantaneous estimates of amplitude and frequency, respectively.  

Despite its simplicity, the ESA accuracy degrades when envelope or frequency variations are fast or when multiple components overlap, as these conditions violate the narrow-band AM--FM assumptions underlying the classical ESA.  In power systems, signals may also depart from simple phasor or cosine models during nonstationary operation \cite{Dervisakdic2020_Beyond_Phasors}.

\vspace{-2mm}

\subsection{Extension to complex signals}

While ESA is formulated for real, narrow-band signals, voltage and current phasors in power systems are inherently complex-valued and exhibit significant magnitude variations during transient operation.
To address these limitations, the TKEO can be generalized to complex signals by employing the symmetric bilinear form:
\begin{align}
\teoB(\bar{x},\bar{y})
  &= \tfrac{1}{2}({\bar{x}}'^*{\bar{y}'} + {\bar{x}}'{\bar{y}}'^*) \\ &
   - \tfrac{1}{4}\!\left(\bar{x}\,{\bar{y}}''^* + \bar{x}^*{\bar{y}}''
   + \bar{y}\,{\bar{x}}''^* + \bar{y}^*{\bar{x}''}\right)\,. \label{eq:psiB}
   \end{align}
   In the special case where $x\equiv y$, this reduces to:
   \begin{equation}
\teoC(\bar{x})
  \equiv \teoB(\bar{x},\bar{x})
   = |{\bar{x}}'|^2 - \text{Re} (\bar{x}\,\bar{x}''^*)\,.
\label{eq:psiC}
\end{equation}
The operator $\teoC(\bar{x})$ is real, additive on Cartesian components, and invariant to constant phase shifts.  
It therefore extends the real TKEO to complex trajectories without imposing narrow-band or slow-envelope constraints.

Alternatively, from \eqref{eq:psiB}, expressing $\bar{x}$ in terms of its real and imaginary parts, one can easily verify that:
\begin{equation}
\begin{aligned}
    \teoC(\bar{x}) \equiv \teoB(\bar{x},\bar{x})
    = \teoR(\text{Re}(\bar{x})) + \teoR(\text{Im}(\bar{x}))\,,
\end{aligned}
\label{complex=real+imaginary}
\end{equation}
as proposed in \cite{hamila_teager_1999}.  This relation allows us to use the identity in \eqref{eq:TKEO-discrete} to calculate $\teoC$ for discrete time.

\vspace{-2.1mm}

\subsection{CTKEO in terms of complex frequency}

Let $\bar{x}=|\bar{x}|e^{\jj\phi}$ denote an analytic AM--FM signal. We recall the complex frequency definition:
\begin{equation}
\bar{\eta}= \frac{\,\,\bar{x}'}{\bar{x}} = \frac{\, \,{|\bar{x}}|'}{|\bar{x}|} + \jj{\phi}'
         = \rho + \jj\omega\,,
\label{eq:beta}
\end{equation}
where $\rho=|\bar{x}|'/|\bar{x}|$ and $\omega={\phi}'$ are the instantaneous magnitude and angular rates, respectively.  
Complex-frequency quantities are related to modal descriptions of oscillatory dynamics used in event analysis; however, in this letter, $\bar{\eta}$ is used as an instantaneous local signal quantity rather than as a global system eigenmode.

Using $\bar{x}' = \bar{\eta} \bar{x}$ and $\bar{x}'' = (\bar{\eta}'+\bar{\eta}^2)\bar{x}$, substitution into~\eqref{eq:psiC} yields:
\begin{align}
\teoC(\bar{x})
  &= |\bar{x}|^2\!\left(|\bar{\eta}|^2 - \text{Re}(\bar{\eta}'^* + \bar{\eta}^{*2})\right) . \label{eq:PsiC-beta}
\end{align}
Expanding $\bar{\eta}=\rho+\jj\omega$ gives $|\bar{\eta}|^2=\rho^2+\omega^2$ and $\text{Re}(\bar{\eta}'^*+\bar{\eta}^{*2})={\rho}'+\rho^2-\omega^2$, leading to the formula:
\begin{equation}
\teoC(\bar{x})=|\bar{x}|^2\big(2\omega^2-{\rho}'\big) \, .
\label{eq:PsiC-final}
\end{equation}

This continuous-time identity follows directly from the complex-frequency representation and does not rely on the classical ESA approximation in \eqref{eq:psi_x}. It assumes, however, that $\bar{x}$ is smooth, nonzero, and that the required derivatives are well defined. The identity separates the local energy into two parts: the rotational component ($2|\bar{x}|^2\omega^2$), and a translation term ($-|\bar{x}|^2\rho'$) that accounts for magnitude variations.

Let us now assume that the voltage trajectory at a node can be represented by two coordinates, say $(v_\alpha,v_\beta)$.  As demonstrated in \cite{MilanoFrenet, Veloso:2026}, this representation is consistent with the zero-torsion case.\footnote{The cases for which the torsion is not negligible are uncommon and have  limited practical interest as they either impact very short periods e.g., few instants following a fault, or arise in stationary conditions characterized by highly unbalanced harmonics, which should be filtered in normal operation.}  This assumption defines the scope of the complex-valued estimator proposed below.  When torsion is non-negligible, the estimator applied to the $(v_\alpha,v_\beta)$ projection should be interpreted as the angular frequency of the projected trajectory, rather than as the complete three-dimensional geometric frequency \cite{Veloso:2026}.  If, in addition to zero torsion, the zero sequence is null, $(v_\alpha,v_\beta)$ corresponds to the coordinates of the conventional Clarke transform \cite{MilanoFrenet}.  Then, we define the complex voltage signal as:
\begin{equation}
    \bar{v} = v_\alpha + \jj v_\beta = |\bar{v}|e^{\jj\theta} \, .
\end{equation}

Assuming $\bar{x} = \bar{v}$ and $\phi=\theta$, and substituting $\rho=|\bar{x}|'/|\bar{x}|$ into~\eqref{eq:PsiC-final} yield:
\begin{equation}
\begin{aligned}
    \frac{\teoC(\bar v)}{|\bar{v}|^2} = 2\omega^2 - {\rho}' = 2\omega^2 + \Big(\frac{\,\,|\bar{v}|'}{|\bar{v}|}\Big)^2 - \frac{\,\,\,|\bar{v}|''}{|\bar{v}|}\,.
\end{aligned}
\label{eq:core-identity}
\end{equation}
Solving for $\omega$ yields the proposed expression for the instantaneous frequency:
\begin{equation}
\boxed{\;
\omega_{\text{TV}}
= \sqrt{\frac{1}{2}\!\left[
\frac{\teoC(\bar v)}{|\bar{v}|^2}
- \Big(\frac{\,\,|\bar{v}|'}{|\bar{v}|}\Big)^{\!2}
+ \frac{\,\,\,|\bar{v}|''}{|\bar{v}|}
\right]}
\;}
\label{eq:omega-est}
\end{equation}

The subscript $\text{TV}$ indicates that expression \eqref{eq:omega-est} is obtained considering $|\bar{v}|$ as a time-varying quantity.
At discontinuities, such as switching or fault clearing, the local derivatives required by \eqref{eq:omega-est} are not well defined in the classical sense. Therefore, the estimator is intended for smooth portions of nonstationary trajectories and should be interpreted with caution at discontinuities.
Unlike the approximate ESA results, \eqref{eq:omega-est} retains the magnitude-dynamics correction term and is therefore bias-corrected with respect to the time-invariant-magnitude approximation. The exactness of \eqref{eq:omega-est} holds for smooth, nonzero complex voltage trajectories satisfying the two-dimensional representation assumed above. In sampled measurements, the accuracy of the estimator depends on the sampling rate, derivative approximation, filtering, and measurement noise.

If $|v|$ is constant or shows slow variations, this expression reduces to $\teoC(\bar{v})=2|\bar{v}|^2\omega^2$, and the instantaneous frequency estimator becomes:
\begin{equation}
\boxed{
\omega_{\text{TI}} = \sqrt{\frac{1}{2} \, \frac{\teoC(\bar{v})}{|\bar{v}|^2}} }
\label{eq:IF-classic}
\end{equation}
 where the subscript $\mathrm{TI}$ indicates that expression \eqref{eq:IF-classic} is obtained by considering $|\bar{v}|$ as time invariant. 
 Equivalently, $\omega_{\text{TI}}$ is obtained from \eqref{eq:omega-est} by neglecting the magnitude-dynamics correction terms involving $|\bar v|'$ and $|\bar v|''$. Thus, $\omega_{\text{TI}}$ is a computationally simpler approximation because it avoids the explicit derivatives of $|\bar v|$. This makes it less sensitive to measurement noise than $\omega_{\text{TV}}$, but it can become inaccurate if voltage-magnitude dynamics are significant.  From \eqref{eq:core-identity}, the simplified estimator satisfies $\omega_{\mathrm{TI}}^2=\omega^2-\rho'/2$; hence, its bias depends on the size of the logarithmic magnitude rate, $\rho'$.

\vspace{-1.5mm}

\subsection{Illustrative example}
To illustrate how neglecting amplitude dynamics can bias frequency estimation, consider the following  signal:
\begin{equation*}
    \bar\upsilon=|\upsilon|e^{\jj\theta}=V(1+V_m\cos(\omega_mt+\phi_m))e^{\jj(\omega_0t+\phi)}\,,
\end{equation*}
which models an amplitude modulation.  The complex frequency components are:
\begin{equation*}
    \omega=\theta'=\omega_0 \quad \text{and} \quad \rho=\dfrac{\,\,|\bar \upsilon|'}{|\bar \upsilon|}=\dfrac{-V_m\omega_m\sin(\omega_mt+\phi_m)}{1+V_m\cos(\omega_mt+\phi_m)}\,.
\end{equation*}
Using the known value, $\omega=\omega_0$, as the reference, the results of Root Mean Square Error (RMSE), Maximum Error (ME), Mean Bias Error (MBE) and Standard Deviation (SD) for both $\omega_{\text{TI}}$ and $\omega_{\text{TV}}$ approaches, and for different sampling frequencies ($f_s$), are shown in Table \ref{tab:table1}. The results consider: $V=\sqrt2\,\,$pu, $V_m=0.1\,\,$pu,  $\omega_0=2\pi50\,\,$rad/s, $\omega_m=0.3\,\omega_0$, $\phi=0.13\,\,$rad and $\phi_m=0\,\,$rad.

\begin{table}[!t]
\scriptsize
\centering
\caption{\\Comparison of $\omega_\text{TI}$ and $\omega_\text{TV}$ estimation metrics} 
\label{tab:table1}
\resizebox{\columnwidth}{!}{%
\begin{tabular}{l|ccc|ccc}
\cline{1-7}
                                               & \multicolumn{3}{c|}{$\omega_{\text{TI}}$}                                                         & \multicolumn{3}{c}{$\omega_{\text{TV}}$}                                                         \\ 
\multicolumn{1}{l|}{{$f_s$ {[}kHz{]}}} & \multicolumn{1}{c}{$10$}    & \multicolumn{1}{c}{$100$}                & $10000$                & \multicolumn{1}{c}{$10$}    & \multicolumn{1}{c}{$100$}                 & $10000$               \\ \hline 
\multicolumn{1}{l|}{{RMSE {[}pu{]}}} & \multicolumn{1}{c}{$0.55$}  & \multicolumn{1}{c}{$0.50$}               & $0.50$                 & \multicolumn{1}{c}{$0.22$}  & \multicolumn{1}{c}{$0.02$}                & $0.00$   \\ 
\multicolumn{1}{l|}{{ME {[}\%{]}}}     & \multicolumn{1}{c}{$0.21$}  & \multicolumn{1}{c}{$0.20$}               & $0.20$                 & \multicolumn{1}{c}{$0.09$}  & \multicolumn{1}{c}{$0.10$}                & $0.00$  \\ 
\multicolumn{1}{l|}{{MBE {[}p.u{]}}}   & \multicolumn{1}{c}{$-0.52$} & \multicolumn{1}{c}{$0.00$} & $0.00$ & \multicolumn{1}{c}{$-0.52$} & \multicolumn{1}{c}{$0.00$} & $0.00$ \\ 
\multicolumn{1}{l|}{{SD {[}pu{]}}}   & \multicolumn{1}{c}{$0.55$}  & \multicolumn{1}{c}{$0.50$}               & $0.50$                 & \multicolumn{1}{c}{$0.22$}  & \multicolumn{1}{c}{$0.02$}                & $0.00$  \\ \hline
\end{tabular}
}\vspace{-3mm}
\end{table}

\section{Case Studies}
\label{sec:case study}

This section evaluates the performance and practical applicability of the proposed frequency estimation, $\omega_{\text{TV}}$, and the approximation, $\omega_{\text{TI}}$.  Two case studies are presented: (i) based on a simulated test; and (ii) based on actual measurement data. $\teoC$ is computed with the discrete expressions \eqref{eq:TKEO-discrete} and \eqref{complex=real+imaginary}, and all the frequency estimations were passed through a discrete Butterworth filter  with cutoff frequency 195 Hz.

The reference frequency used in the comparisons is the geometric angular-frequency estimate $\omega_v$, computed from the voltage trajectory in the $(v_\alpha,v_\beta)$ plane.  This benchmark is selected because it is consistent with the complex-frequency and geometric interpretations adopted in this work.  Moreover the consistency of the geometric frequency with conventional estimators, e.g., PLLs and synchrophasors has been established in previous works, e.g., \cite{Veloso:2026}.  In the field-measurement case, $\omega_v$ should be interpreted as a computed reference estimator rather than as an exact ground-truth frequency.

For a discrete sinusoidal component with digital angular frequency $\Omega=\omega/f_s$, the discrete TKEO response scales with $\sin(\Omega)$ rather than $\Omega$. Hence, $\sin(\Omega)=\Omega-\Omega^3/6+O(\Omega^5)$, and the sampling-related error decreases as the sampling rate increases. This effect is only associated with the sampled implementation and does not modify the continuous-time identity derived in Section \ref{sec:ESA}.  Besides this sampling effect, when implemented in online applications, the value of $\teoC(\bar{\upsilon}_n)$ is computed when the subsequent sample $\bar{\upsilon}_{n+1}$ is available, introducing a one-sample delay.

\vspace{-1mm}

\subsection{EMT Simulation of IEEE 39-Bus System }

\begin{figure}[th]
  \centering
  \includegraphics[scale=0.49]{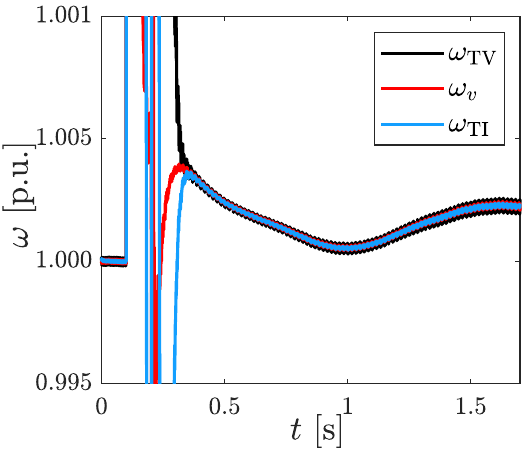}
  \includegraphics[scale=0.49]{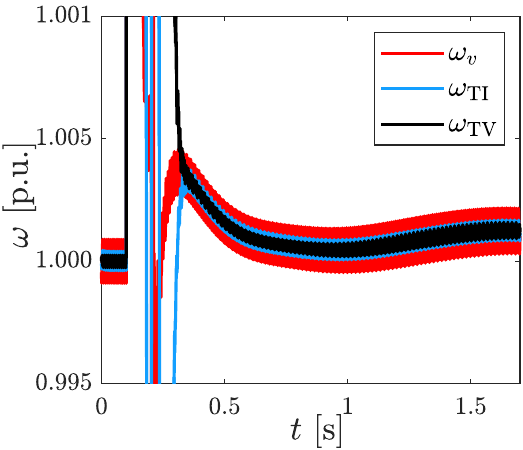}
  \caption{Frequency results bus $26$, of the IEEE 39-bus system for a secured three-phase fault under: (left) balanced conditions with Gaussian noise in the voltage; and (right) unbalanced operation.}
  \label{fig:39bus}
\end{figure}

In this case study, the IEEE 39-bus system, from DIgSILENT PowerFactory software tool, is assessed. 
The considered signal is the three-phase voltage at bus $26$ following a three-phase fault at bus $4$ at $t = 0.2$ s, cleared at $t = 0.3$ s.  The tests consider two different scenarios: balanced system operation with Gaussian noise in the voltage measurement and unbalanced operation, see Fig.~\ref{fig:39bus}.

Results under balanced conditions with noise show that estimations from the three approaches do not differ significantly.  Both scenarios show that frequency estimation from $\omega_{\text{TI}}$ and $\omega_{\text{TV}}$ closely match because the variations in the voltage magnitude are relatively slow and small, and are mitigated by filtering the time series.  On the other hand, in the unbalanced scenario, $\omega_{\text{TI}}$ and, especially, $\omega_{\text{TV}}$ show lower oscillations, before and after the fault, with respect to $\omega_v$, as this quantity captures the imbalances as a distortion in the voltage curvature; indicating that the proposed estimators are less affected by the curvature distortion introduced by the unbalanced condition in this case.

\subsection{Voltage Measurements at a PV Power Plant}

We consider field measurements of the voltage at the point of connection to the grid of a PV power plant in Moralejo, Spain.  Accurate computation of local derivative-based estimators under noisy measurements requires filtering; in this case, $\omega_{\text{TV}}$ requires filtering both $v'$ and $v''$, whereas $\omega_v$ only  requires filtering $v'$.

\begin{figure}[th]
  \centering
  \includegraphics[scale=0.8]{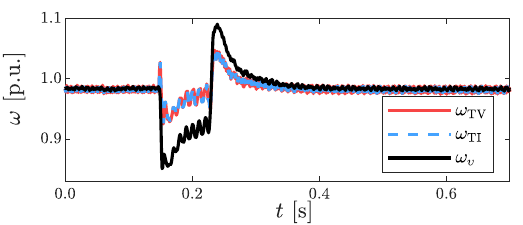}
  \caption{Frequency estimation results from real voltage measurements at the point of connection to the grid of a PV power plant.}
  \label{pv}
\end{figure}

Results in Fig.~\ref{pv} show that $\omega_{\text{TI}}$ and $\omega_{\text{TV}}$ follow the dominant trend of $\omega_v$, although visible discrepancies appear during intervals affected by measurement noise, filtering, and derivative sensitivity. These discrepancies are expected because $\omega_{\text{TV}}$ requires higher-order derivative information, whereas $\omega_v$ only requires first-order derivatives.
Although $\omega_{\text{TI}}$ does not account explicitly for magnitude variations, the combined effect of voltage filtering and the inherently slow dynamics of $\rho$ under realistic operating conditions makes this approximation practically very close to the full time-varying estimate.

Thus, $\omega_{\text{TI}}$ provides a reliable and computationally simpler approximation of instantaneous frequency, whereas $\omega_{\text{TV}}$ yields a more precise estimate when the correction terms associated with magnitude dynamics become relevant. Compared with $\omega_v$, the proposed approach also provides a decomposition of frequency estimation into interpretable terms: the dominant contribution is given by $\teoC(\bar v)/|\bar{v}|^2$, while the magnitude-rate terms act as corrections that determine the accuracy of the simplified expression $\omega_{\text{TI}}$.

\section{Conclusions}
\label{sec:conclusion}

This letter presents a formulation of instantaneous frequency derived from the CTKEO. Starting from the dynamic-signal identity and applying it to a complex voltage trajectory, we obtain a local expression for the angular frequency that explicitly retains the correction associated with voltage-magnitude dynamics. The resulting identity is exact in continuous time for smooth, nonzero two-dimensional voltage trajectories, whereas its sampled implementation depends on the derivative approximation, filtering, sampling rate, and measurement noise.

Future work will address a full three-dimensional extension for nonzero-torsion trajectories and a broader validation against PMU, PLL, Hilbert, and adaptive-filtering-based frequency estimators under diverse disturbances.  PMU-based event detection will also be investigated as a potential application of the proposed method.


\vfill
\end{document}